# Strategy-proof Pricing Approach for Cloud Market

*Chetan Chawla, **Inderveer Chana

*chetanchawla16@gmail.com, **inderveer@thapar.edu

Computer Science and Engineering Department, Thapar University, Patiala, India-147004

*Abstract*— In this paper, we design and develop a pricing model applicable to strategy proof pricing. To provide an economic stability towards its consumers. The economic model we use is Vickrey-Clarke-Groves (VCG). By this each service provider has to provide a true cost of its services in the cloud market. For the selection of suitable service for the consumer we adopt a dynamic programing based algorithm and VCG is used to calculate the payment. Strategy proof pricing offers a unique cloud pricing service that takes the complexity out of traditional pricing and enables cloud providers to price accurately, consistently and competitively.

## I. Introduction

Everyone in this universe needs a substantial alteration of its working area. As the time passes quickly, within seven decades from the making of the very first digital computer, ENIAC [1], we are breathing in the age of the Internet, and we cannot envisage our life with no Internet. The internet provides its consumers many facilities (services) such as Emails, Financial, Banking Solutions, Insurance, Social Networking, Stock exchange and E-Business etc. By the existence of this emerging technology different researcher sees Cloud Computing is one of new area to research and serve well in this area. The cloud computing practice is an on-demand/requested services of package supply generally on the basis of 'pay_as_you_go' model. There are many explanations of cloud computing, which is stated by different authors in their research area and we will consider the very next statement/definitions stated below.

Buyya Rajkumar has defined it as follows: "Cloud environment is a type of parallel and distributed system consists of a gathering of virtualized machine over channel and interlinked computers that are dynamically subscribed provisioned services. It also provides resources based on Service Level Agreements (SLA) recognized between the service provider and consumers" [2].

As we discussed prior that different researcher/authors from different background provide different definitions for this emerging field Cloud Computing, so we add yet another meaning for this emerging research oriented area.

"It is a large scale parallel and distributed computing archit-cture in which virtualization, multi-tenant, scalable, managed computing power, memory, infrastructure, software, platform are delivered on demand to external customer in a business oriented environment".

"Cloud Computing" approach used in daily life is an on-demand services provider to its users when they demand it over an internet; it uses several traditional computing as well as new networking and computing approaches. A cloud computing is a way to interaction between its providers and its consumers over a channel and on to a cloud compositions. It holds techniques these are Vz's (Virtualizations), computing with distributed procedure, resource provisioning etc. The QoS plays a vital role in the cloud compositions i.e. market which are high scalability, reliability, testability, maintainability. Cloud compositions interface should evade fault tolerances and optimal pricing solutions to its consumers a well increase in its revenue.

From the demand of several consumers to provision all cloud services which leads to development of cloud compositions. The project aims to provide services to its users on the basis of class features like price and quality. The traditional model uses an economic model which is reverse auction and in this first auction leads to several unwanted outcomes for its modules users or providers as well. To overcome these unwanted outcomes VCG mechanism is used for the price attribute and for quality traditional iterative model is enhanced. By VCG each providers has to suggest their true cost of the resources over this environment of compositions. The computing model can be done by a dynamic programing algorithm for the selection of best services over the cloud compositions featured.

## II. Related Work

Cloud computing, mainly attentions on two trends from the puddle of information technology – i) IT Efficiency and ii) Swiftness of Business, from the areas of IT effectiveness, this approach should utilize the pricing techniques for contact establishment between resources provider and its end side consumers. Area of distributing computing provides its stakeholders to explore their business perspective, it provides them the thought in their business and it is done by the actual owner of these contract no need of any third party stakeholder [3]. From the business side point of view, there are many hurdles in this ever changing environment which is cloud computing. Economic aspect is a key factor in this area and pricing model technique is very crucial in the generating revenue of stakeholder of various cloud service providers.

In this demanding world, to exist in this digital world by any organization, then it has to provide very flexibility in pricing, rapid provisioning of resources, green solutions with low carbon footprint, scalability, testability and stability over old-style IT deployment models.

Often used terminologies in cloud computing are defined as: Utility is defined as the collection of several resources as a metered service; resources can be natural or artificial. Utility computing deals with several resources of computing environment that can be storage, processing power, network throughput, etc. In this kind of model users can request the service they needed and consumers are charged by the amount of their usages of that service and have to pay that to service providers. Utility computing has this best feature which is no earlier investment of the resources. In the vision of utility computing several paradigms of computing take a step towards this to provide a utility computing this includes Grid Computing, Cluster computing and newly Cloud Computing. Utility Computing is not a new approach in 1969, Leonard Kleinrock [4] the scientists of Advanced Research Projects Agency Network (ARPANET) said, 'Today's network for computers are still in their beginning age, but as they mature and become cultured, we will possibly see the spread of 'computer utilities' which, like present electric, gas, water and cellular utilities, will service individual homes and offices through the world wide'. Cloud Computing will be essential in terms of service provider utilities to end user in the 21st century.

With the ease of internet we can access any information from anywhere within a few seconds, this technology consists of data centers, their work is to provide content to requested end user accurately and consistently. Cloud computing uses this approach to establish its architecture and in this it provides business application capabilities as a service to end consumer that can be accessed over internet. This technology is adapted by many big giants such as Amazon, Google, Microsoft, IBM and Sun Microsystem. They have started to build new data centers around the globe to provide reliability and availability in case of failure of any data enter. Cloud service providers charge their consumers with 'Pay-as-you-go' Model by which the consumer has to pay for the services he/she used. Providers can make their profits by charging their customers for the usage of the service which they provide over the internet. Cloud computing is an on demand service providing technique, so consumers should have guaranteed that the services they are paying for should deliver to them uninterrupted. This is provided through SLAs (Service Level Agreements) between the consumers and providers.

In [5] Buyya et al. discuss various models for various resources (goods and services) for grid computing with economics perspective. Following Table I shows the complete perspective of these models.

TABLE I: SUMMARY OF RELATED WORK [15-16]

| Economic Model for Resources | Narration |
|---|---|
| Bargaining Model | This class of models represents those models in which resource (Grid or Cloud) providers and resource consumers discuss with each other until they meet their aims and agree on a price. |
| Commodity Market Model | In this model, various pricing model can be used, the approach of this model is 'more you consume more you have to pay'. The price of resources is calculated on the basis of consumption. Different pricing model which comes under this class are- Flat fee, Subscription, Usage Duration. |
| Posted Price Model | The model is similar to the commodity market model but the additional paradigm this model provides is known as advertising. Providers attracts their consumers through advertisements. |
| Auction Model | Auctioneer set the rules which is accepted by both providers and consumers. Different consumers auctioned for the resources and negotiate and agreed on a price. |
| Monopoly Oligopoly | This model shows the dominance of a Grid Service Provider i.e. GSP. GSP is the only provider of particular resource, service or protocols. User have to restrict to the price set by the GSP.- |
| Contract-Net model | In order to control the usage of services and resources, model is used for contracting between providers and consumers. |
| Bartering Model | This model works on the perspective hybrid cloud, individual or institutions create a computing environment and share their resources. |
| Bid Based Model | This model works on a bid value or a threshold value, which states the amount of share customer had for various resources. |

The pricing technique simulated by Xiangang Zhao et al. [6], in based on demand prediction and task classification. As stochastic process with a specific property know as Markov Chain, by which the authors are able to predict the demand of resources. As grids are dynamic in nature so as the available resources in the grid market and they are also stochastic in nature so we can easily apply the Markov Chain. By the help of Markov Chain we predict the demand/load of the given resources in the future and then a mechanism to adjust the prices of the resources. As price and demand are mutually dependent to each other so we have to apply pricing techniques which consider in the profit of resource providers as well as load balancing of the available resources. They also define the tasks which are exclusive task and shared task depending upon this they provide two different pricing strategies.

### III. MODEL FORMATION

This project overcomes the disadvantages associated with first price auction. Because when first price auction took place, cheaper price and effective cloud service will be selected. This first price auction some unwanted results to both consumers and service providers. The solution for this is the VCG mechanism which is Vickrey-Clarke-Groves, by this each service provider has to provide a true cost of its services in the cloud market. For the selection of suitable service for the consumer we adopt a dynamic programing based algorithm and VCG is used to calculate the payment. The proposed algorithm for this is an adaptation of Masahiro et al. [7] which solve the selection of service problem in quasi-polynomial time. This system's feasibility depends upon the experiments conduct by [7].

Strategy proof pricing offers a unique cloud pricing service that takes the complexity out of traditional pricing and enables cloud providers to price accurately, consistently and competitively. This framework delivers a comprehensive solution for convergent pricing across cloud services, helping providers to achieve the full promise of cloud computing. The automation model reduces the cloud vendor lock in, efforts, human intervention, and improves customer satisfaction, service quality, scalability and control on investments. This model offers an uninterrupted service with payment collection of payment using third party component. Strategy proof pricing optimizes cloud pricing processes, enabling cloud providers to competitively monetize their offerings and increase profits. Table II shows the benefits of our proposed model. The key features of this service include:

- High flexibility, including the ability to price the cloud computing resources, support for sophisticated price-discounting, and support for both customer, provider, admin, budget manager and partner account balance types
- Definition of cloud products, services and pricing models by SaaS cloud provider.
- Strong systems integration capabilities and support.

TABLE II: Comparison of Strategy proof pricing with traditional methods [8-14]

| Criteria | Strategy Proof Pricing | Traditional Methods |
|---|---|---|
| Methodology | Adaptive | Predictive |
| Interaction | More | Less |
| Response | Rapid process | A slow process |
| Plan selection | Reverse Auction | No |
| Policy Plan | VCG Model | Fixed or sometimes Pay Per Use |
| Feedback | Yes | No |
| Quality aspects | Yes (Usability, Security etc.) | No |
| Maintain records | Yes (DBMS) | No |
| Payment | By third Party Component | By banks |
| Flexible & reliable | Yes | No |
| Consumption | Transparent | No |
| Report generation | Automatic | Manually |
| Budget Management | Yes | No |
| Compare cost | Yes | No |
| Accuracy | More | Less |

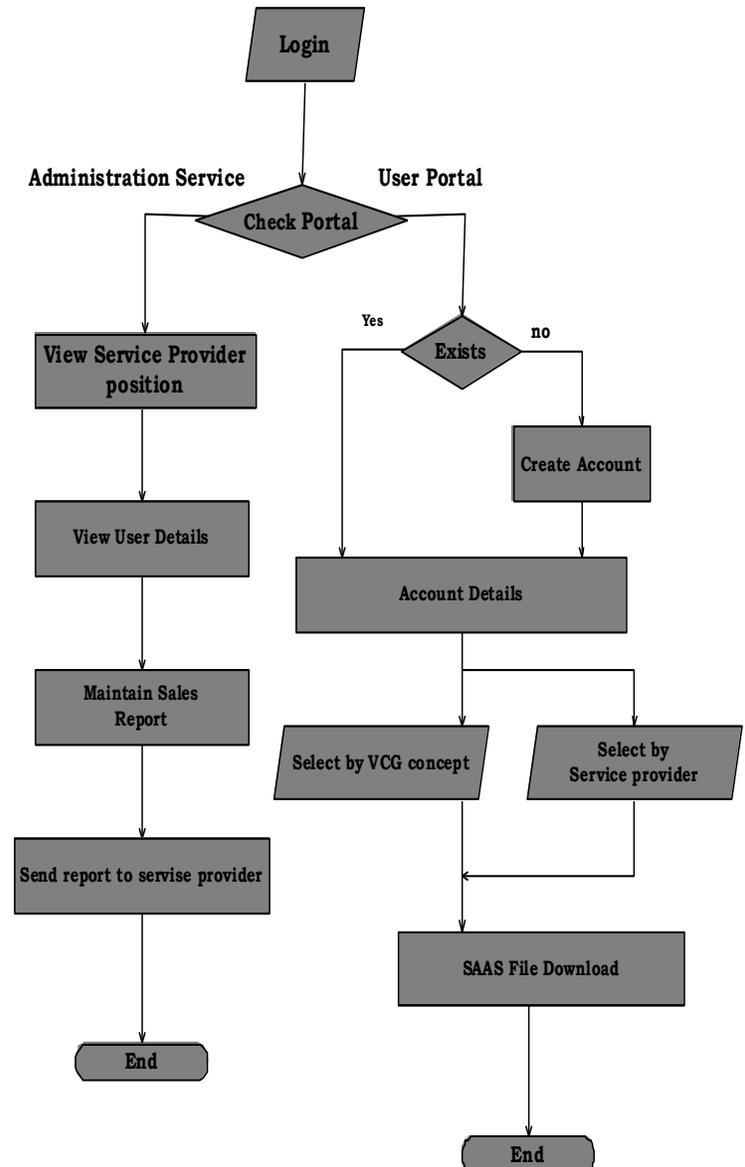

Fig. 1 Flow Graph of strategy proof pricing project

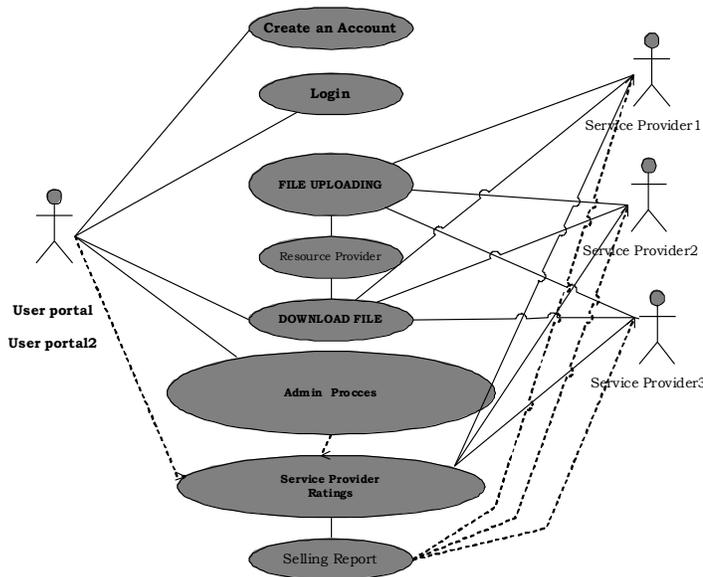

Figure 2. Use Case Diagram of strategy proof pricing project

Figure 1. Shows the flow of strategy flow pricing technique used by VCG. VGC economic model helps formulation of VCG efficiently. Figure 2. Shows the use case of our proposed model is give the unified model of our architecture.

## IV. CONCLUSION

The minor report that we discussed here focus on the introduction of cloud computing and need of a distributed cloud composition services. We also discusses the various technique which are used and implemented over various cloud markets and some simulated studies related to cloud services over cloud marketplace. The project manages both the price and SaaS perspective user to provide a better services to its consumers over a distributed cloud compositions. The work conducted under the minor project gives a technique to overcome the problem finding a suitable match of service to its consumer with factors such as price and quality. A cloud composition module is design to implement the project how it works over a distributed environment on the basis of price and quality. The project is developed in the .Net architecture and run on local host. The goal of this service is to maximize the efficiency of cloud compositions. Purpose is to focus on the selection of iterative approach to find a service provider and VCG for improvement of traditional services.

## V. FUTURE SCOPE

- This work shows how SLA violations can be overcome between the service providers and its consumers using a distributed cloud composition environment.
- This work can be deployed to Apache Cloudstack [20]. An open source cloud IaaS provider. Using this we can build our own cloud infrastructure as a service and can deploy this pricing techniques to its consumers.
- Several quality of service parameters can be used to deploy this project over a cloud infrastructure.
- Statistical technique to generate the revenue of provider can also be used like Time Series, ANOVA, and Markov Chain etc.
- Rewards can be added to pricing strategy to stick with the consumers which uses the services from the past and attract the new consumers.
- Service selection model can be enhanced more towards iterative search of service with the VCG mechanisms.